# On the Need of Preserving Order of Data When Validating Within-Project Defect Classifiers

Davide Falessi[a]*, Jacky Huang[b], Likhita Narayana[b], Jennifer Fong Thai[b], Burak Turhan[cd]

[a]University of Rome "Tor Vergata", Italy

[b]California Polytechnic State University, CA, USA

[c]Monash University, Melbourne, Australia

[d]University of Oulu, Finalnd

davide.falessi@uniroma2.it {jhuang81, lnarayan, jfthai}@calpoly.edu, burak.turhan@monash.edu

*\* = corresponding author*

**Abstract.** [Context] We are in the shoes of a practitioner who uses previous project releases' data to predict which classes of the current release are defect-prone. In this scenario, the practitioner would like to use the most accurate classifier among the many available ones. A validation technique, hereinafter "technique", defines how to measure the prediction accuracy of a classifier. Several previous research efforts analyzed several techniques. However, no previous study compared validation techniques in the within-project across-release class-level context or considered techniques that preserve the order of data. [Aim] In this paper, we investigate which technique recommends the most accurate classifier. [Method] We use the last release of a project as the ground truth to evaluate the classifier's accuracy and hence the ability of a technique to recommend an accurate classifier. We consider nine classifiers, two industry and 13 open projects, and three validation techniques: namely 10-fold cross-validation (i.e., the most used technique), bootstrap (i.e., the recommended technique), and walk-forward (i.e., a technique preserving the order of data). [Results] Our results show that: 1) classifiers differ in accuracy in all datasets regardless of their entity per value, 2) walk-forward outperforms both 10-fold cross-validation and bootstrap statistically in all three accuracy metrics: AUC of the selected classifier, bias and absolute bias, 3) surprisingly, all techniques resulted to be more prone to overestimate than to underestimate the performances of classifiers, and 3) the defect rate resulted in changing between the second and first half in both industry projects and 83% of open-source datasets. [Conclusions] This study recommends the use of techniques that preserve the order of data such as walk-forward over 10-fold cross-validation and bootstrap in the within-project across-release class-level context given the above empirical results and that walk-forward is by nature more simple, inexpensive, and stable than the other two techniques.

# 1 Introduction

As testing remains one of the most important activities in software engineering, prioritizing test cases by predicting components likely to be defective is vital to prioritizing effort allocation. The software engineering community has provided significant advances in classifiers, and more advances are probably on their way. Prediction models can support test resource allocation by predicting the existence of defects[1] in a software module (e.g., class). Specifically, classifiers aim to estimate a categorical variable, i.e., the existence or lack of at least one defect in a software module, aka class.

In this paper, we are in the shoes of a practitioner who uses previous project releases' data to predict which classes of the current release are defect-prone. In this scenario, the practitioner would like to know which classifier to use, i.e., which classifier will provide the most accurate prediction, and needs a validation technique to make a decision. Such a scenario is similar to the one of a researcher working on improving the classifier's accuracy using noise removal [1][2][3], tuning [4][5][6], rebalancing [7][8], and feature selection [9] technologies. This researcher wants to investigate if the technology improves the performance that a classifier will have in the future. Thus, in this paper, the term classifier refers to a way to make a binary prediction, and hence it includes all possible related technologies such as noise removal, tuning, and feature selection. In other words, the same classifier with different configurations can be seen as two different classifiers with and without the use of a technology, e.g., tuning. Thus, the scenario of choosing a classifier among many is equivalent to the scenario of validating the use of these technologies.

A validation technique, hereinafter "technique", defines how to measure the prediction accuracy of a classifier by prescribing a specific way to construct one or multiple sets of data where the classifier is first trained and then tested. This accuracy can be used for making informed decisions about the classifier to use in the future or, similarly, to validate technologies aimed at improving classifiers' accuracy. In this paper, we investigate which validation technique recommends the most accurate classifier. Specifically, we investigate the following research questions:

- *RQ1: Do classifiers vary in accuracy?* Before comparing techniques we check if there is any reason to use techniques or if at the contrary, there is not much difference among their accuracies and hence we can randomly choose the classifier. Entity Per Value (EPV) [10] is defined and computed as the number of occurrences of the least frequent class, i.e., defective classes, divided by the number of metrics used by the classifier, e.g., number of developers. According to previous studies, classifiers differ in accuracy only in datasets with low Entity EPV [11] [12]. However, no previous study investigated the impact of EPV on classifiers accuracy in the within-project multi-release context.

- *RQ2: Do techniques vary in accuracy?* Our experience in the industrial context [13][14][6] shows that one relevant and practical problem when institutionalizing a classifier is the selection of the specific classifier to use. This problem does also apply when selecting a specific configuration of a classifier, i.e., tuning [4][5][6], since different configurations of the same classifier can be seen as different classifiers, e.g., different instances of the same abstract classifier. No previous study investigated the impact of validation techniques in recommending classifiers in the within-project multi-release context.

---

[1] As in Hall et al. [31] we use the term defect to indicate a fault or a bug. A failure is a possible result of a fault occurrence.

In our methodology we use the last release of a project as the ground truth to evaluate the classifier accuracy and hence the technique's ability to recommend an accurate classifier. We consider nine classifiers, 13 open and two closed projects, and three validation techniques: namely 10-fold cross-validation (i.e., the most used technique), bootstrap (i.e., the recommended technique [11]), and walk-forward (i.e., a technique preserving the order of data).

Our results show that:

1. In contrast to previous studies [11] [12], classifiers differ in accuracy also in datasets with high EPV. Thus, *it is important to choose the classifier by using a technique*.

2. Regarding which validation technique to use, *walk-forward outperforms both 10-fold cross-validation and bootstrap in recommending accurate classifiers*.

3. The defect rates change between the second and first half of the projects investigated, potentially explaining the observation that *validation techniques that do not preserve the order of data (i.e., 10-fold cross-validation and bootstrap) provide poorly realistic results*.

The remainder of the paper is structured as follows. Section 2 reports on related work. Section 3, 4 and 5 describe the design, results, and conclusions. The threats to validity are discussed in Section 6. Section 7 concludes the paper.

## 2  Related work

In this section, we first provide an overall view of time-series vs. non-time-series validation techniques, then report the specific validation techniques used in defect prediction research during the last decade, and finally conclude with a discussion of other methodology papers that focused on the impact of different validation techniques to differentiate our work from earlier studies.

### 2.1  Validation Techniques: Time-series vs. Non-time-series

Time-series techniques are used in statistics, signal processing, pattern recognition, econometrics, mathematical finance, weather forecasting, earthquake prediction, electroencephalography, control engineering, astronomy, communications engineering, and largely in any domain of applied science and engineering which involves temporal measurements [17][18]. Walk-forward has been largely used in for validating models that predict stock prices [19][20][21][22]. In walk-forward, the dataset is divided into parts, i.e., the smallest units that can be ordered, e.g., a release of a project. Then, parts are chronologically ordered, and in each run, all data available before the part to predict is used as the training set, and the part to predict is used as test-set. Afterward, the model accuracy is computed as the average among runs. The number of runs is equal to one less than the number of parts. For instance, Figure 1 (a) describes the walk-forward technique; the parts used for training are in blue, the ones used for testing are in green, the ones not used are in white. Figure 1 (a) describes a dataset related to a hypothetical project of five releases, i.e., five parts, and four runs. In the first run, the first part is used as training, and the second as testing, in the second run the first two parts are used as training and the second as testing, and so on. The accuracy is averaged among the four runs.

Non-time-series techniques vary in the way the dataset is split into the train and test sets as it could be due to random sampling, with or without replacement, etc. The v-fold (aka k-fold) cross-validation is the most used non-time-series technique, and it makes use of random sampling strategies to construct several training and test sets on which the accuracy of the model is averaged [11] [23]. In a k-fold cross-validation setting, the model accuracy is the average among runs and the number of runs is equal to the number of folds, i.e., parts, in which the data is divided. Figure 1 (b) describes a five-fold cross-validation scenario. In Figure 1 (b), the dataset is randomly divided into five parts of equal size. Since the parts, i.e., folds, are randomly generated, to minimize effects related to random sampling, the procedure is usually repeated several times. Empirical studies recommend that v-fold cross-validation works well if "v" and repetitions are ten [16]. We note that when "v" equals the number of observations, then the technique is called leave-one-out. As we can see in Figure 1 (b), in all runs other than the fifth, future data is used as a training set. An option of the v-fold cross validation is stratification, i.e., enforcing the parts to have the same defect ratio. The advantage of this stratified approach is to support model training as the model can analyze heterogeneous data during training. The disadvantage is that it reduces realism and overestimates model accuracy as it assumes the test set to have the same defect ratio as the training set.

The bootstrap method, proposed by Efron and Tibshirani [24], consists of creating the training set by randomly sampling data with replacement and using the original set as the test set [25]. Several variants of bootstrap exist including the optimism reduced, where the effect of testing on the same data used as training is removed by subtracting, to the measured accuracy, the accuracy of the model tested on the training set. A further variant is called out-of-sample bootstrap where the model instead of being tested on the original dataset is tested on the data of the original dataset that is not used (sampled) in the training set. Tantithamthavorn et al. recently recommended the use of this out-of-sample bootstrap to measure the accuracy of defect classifiers [11]. Figure 1 (c) shows an example of out-of-sample bootstrap where in the first run the parts selected as training are the 2nd, 4th and 5th, and the remaining parts (i.e., first and third) are used as test-set.

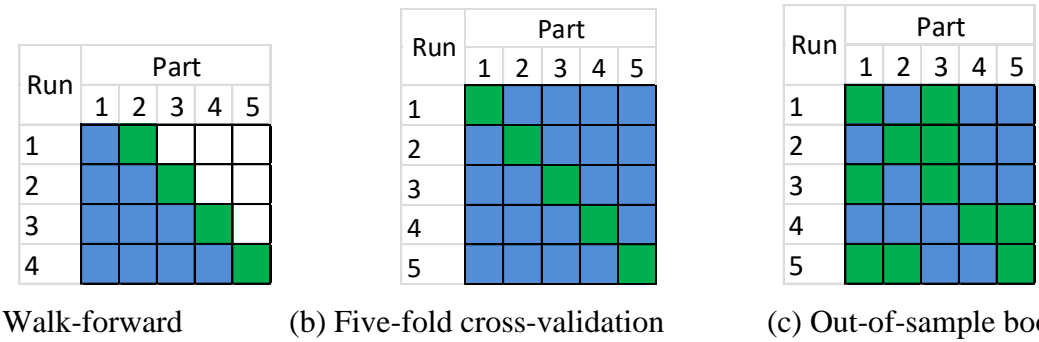

(a) Walk-forward     (b) Five-fold cross-validation     (c) Out-of-sample bootstrap

Figure 1: Different validation techniques; the parts used for training are in blue, the ones used for testing are in green, the ones not used are in white.

We refer to Bergmeir and Benítez [15] for a detailed discussion of time-series techniques and Kim [26] on non-time-series techniques.

## 2.2 Which Validation Techniques Are Used in Defect Prediction?

To have an overall view about which techniques have been used, why they have been used, and whether they preserved temporal order of data, we performed a non-committal literature review [27]. Please note that our aim here is not to provide a systematic literature review of related literature, which is out of the scope of this paper, but rather to get insight regarding the use of time-series validation techniques in defect prediction research during the last decade.

To identify relevant studies, we performed a search on Google Scholar using the following search string in the title: *bug\* OR defect\* OR fault\*) AND (predict\* OR estimate\*).* This string allowed us to find studies related to defect prediction using different synonyms and declinations such as bug or defects and prediction or estimations. As in a previous study [28], we chose Google Scholar because it provides the complete coverage of scientific literature; moreover, it does not suffer from the idiosyncrasies reported earlier for such collections of scientific papers as IEEE Xplore and ACM DL [29].

The search resulted in identifying 9,763 studies. We filtered out studies older than ten years and published in venues other than peer-reviewed journals, leading to 106 studies. Then, we removed studies not applying any validation technique (e.g., secondary studies) and where the variable predicted by the model is numeric (i.e., regression studies that predict the number of defects). This led to a final set of 51 studies. To ensure reliability and reduce researcher bias, the gathering and filtering procedure was performed by the third and the fourth author independently. There were two inconsistencies among the third and fourth author, e.g., a Cohen Kappa agreement of 96.2% [30]. The first author resolved these two inconsistencies and cross-checked a random sample of ten studies for correctness, all ten classifications from both authors resulted correct.

To identify which technique has been used in each study, we analyzed the section of the studies discussing the evaluation procedure. We started from an empty list of techniques and we added techniques as we found them in the identified studies. To identify the criteria used for choosing techniques, we checked the reported rationale. Similarly, we started with an empty list of criteria and we added them as we found them in the studies. Finally, we proceeded by checking the types of datasets, i.e. mono-release vs. multi-released, used in the studies and hence the proportion of times whether time-series techniques were feasible and/or not used.

Table 1 reports the proportion of techniques used in past studies. The most notable result is that only one time-series technique is used, i.e., the walk-forward cross-validation, and this technique is used by only 6 (9%) studies. The most used technique is by far k-fold cross-validation with 31 (61%) studies. Additional analysis of data shows that 10-fold cross-validation is used by 26 (51%) studies and hence it is the most used type of k-fold technique and the most used technique in general. This result is in line with Tantithamthavorn et al. [11].

Table 1: Number of studies using a technique.

| k-fold cross validation | Holdout | Bootstrap | Different projects | Time-series cross-validation |
|---|---|---|---|---|
| 31 | 5 | 9 | 2 | 6 |

As for the rationale for using specific validation techniques, the most used criterion (37%) is being "specific to the research context". For instance, a past study aimed at validating a technology related to cross-company prediction used a specific sample of projects as training set and other projects as a testing set. The second most used criterion (29%) is "not reported", specifically, it was impossible for us to retrieve why the technique was used. To ensure that this result was not biased by individual members of our research team, all three researchers independently analyzed the studies with the unknown criterion and unanimously concluded that the criterion was not reported. The third most used criterion (26%) is being "widely used" in the past. The last used criterion is being "used by specific studies" (8%).

## 2.3 The comparative study on validation techniques

This study has been highly inspired by Tantithamthavorn et al. [11] who recently found that the most used technique is 10-fold cross-validation and, by comparing 12 different techniques, they recommend the use of out-of-sample bootstrap technique [23]. We share with them the need to better investigate the intrinsic and practical differences among validation techniques. We built on their contributions by contextualizing it on a specific context where the first author has industrial experience, i.e., the within-project defect prediction. Other than the context, there are several methodological differences between our and the Tantithamthavorn et al. [11] study:

1. We measure the accuracy of the technique as the accuracy of the classifiers it recommends to use.
2. We interpret the technique accuracy by comparing it to the best, worst and medium recommendation.
3. We consider a higher number of classifiers and a smaller number of techniques.
4. We investigate if the conditions of non-time-series techniques are met. This is powerful since independent from the set of classifiers, techniques or metrics.

To the best of our knowledge, no study other than Tantithamthavorn et al. [11] explicitly compared validation techniques, but there are studies that are related to model selection criteria which we discuss in the next subsection.

## 2.4 Studies on model selection criteria

Hall et al. [31] performed a systematic literature review to investigate if the context of models, the independent variables used, and the modeling techniques applied impact the accuracy of classifiers. Their results show that "*the methodology used to build models seems to be influential to predictive performance*" [31]. Their results motivate our work (RQ2) in investigating if the same classifier has different accuracies across different validation methodologies, i.e., the techniques.

Myrtveit et al. [32] demonstrated that the selection of the most accurate classifier is highly impacted by specific choices made during the experimental design such as accuracy metrics. We share with them the need to better investigate how experimental choices impact results. However, they investigated models predicting a numeric variable (i.e., effort) whereas our prediction variable is binary (i.e., class defectiveness). Another major difference is that they focus on evaluating the impact of the accuracy metric on the decision of which classifier is best whereas in our RQ1 we evaluate the impact of the validation technique on the classifier accuracy measurement. We do not investigate the impact of accuracy metrics because previous studies recommend AUC to be a reliable metric, not sensitive to defect ratios, in the context of class defectiveness prediction [33]. In our view, no validation technique or metric is better or worse than another; everything is context dependent. Specifically, the issue is to understand the extent to which current techniques and metrics replicate specific classifier usage scenarios.

Kocaguneli and Menzies [34] recommend the use of leave-one-out technique (i.e., k-fold with k equal to the size of number of observations) for validating models on effort estimation. Their results show that 10-fold cross validation produces similar results and similar time than leave-one-out; however, leave-one-out is perfectly replicable and hence should be preferred. On the other hand, Sigweni et al. [35] argue that using a time based approach is not only replicable and more realistic, but also significantly more conservative than leave-one-out methods in estimating software effort. These results from effort estimation literature also motivate us to study the same concept in the context of defect prediction.

McIntosh and Kamei [36] recently investigate the freshness of data in the context of just-in-time prediction, i.e., the prediction of the defectiveness of a single commit. Their results show that classifiers have a better accuracy if the oldest data, because it is outdated, is not used as input. Despite the high general relevance of the results, our study is different in aim and context. Specifically, in our context the predicted unit is a class rather than a commit, and since our projects are industrial and random rather than rapidly evolving, then our observations are several orders of magnitude smaller than the ones used in McIntosh and Kamei. Further, while their work is targeted at developers to receive 'just-in-time' feedback about their commits, the primary stakeholders of our work are the Quality Assurance teams.

Jonsson et al. [37] investigate the automated bug assignment via machine learning. Their results are in line with McIntosh and Kamei, i.e. classifiers have a better accuracy if the oldest data is not used as input. Still, the number of our observations is much smaller than the ones used in Jonsson et al. [37]. Jonsson et al. results also shows that: 1) ensemble classifiers outperform the use of single classifiers, and 2) cross-validation yields higher prediction accuracy than time-series techniques. Specifically, they suggest complementing cross-validation with a sequestered test set; we followed their suggestion as better explained in Section 3.1.3.

In conclusion, our paper, compared to the state of the art, provides the following contributions:

1. We show that EPV does not impact the classifier accuracy (RQ1). This contrasts with results of previous studies in different contexts.
2. We compare a time-series versus two non-time series techniques (RQ2).
3. We compare validation techniques according to the chosen model accuracy (RQ2).
4. We compare validation techniques according to the bias (not only the absolute bias) (RQ2).
5. We compare the validation technique's performance to three hypothetical techniques (RQ2).
6. We show that the techniques suggested in Tantithamthavorn et al. [11], such as out-of-sample, should not be used because they are more complex, expensive, instable, and inaccurate than walk-forward. Thus, it highlights that no size fits all; the technique to use depends on the conclusions to draw, the property of the available datasets, and the level of realism with the classifier usage scenario (RQ2).

7. We show that walk-forward overestimates classifier accuracy less than the other two techniques, (RQ2).
8. We show that the order of data needs to be preserved since the defect rate significantly changes across releases of the same project (RQ2).

## 3 Design

### 3.1 RQ1: Do classifiers vary in accuracy?

The aim of this research question is to motivate the use of a validation technique to recommend the classifier to use. Thus, in this research question, we investigate if the accuracy, as measured on the same project, varies among classifiers. If this isn't the case, then we could choose the classifier to use randomly rather than using any, likely complicated, validation techniques. Thus, our null hypothesis is

*H01: The accuracy does not vary either among classifiers or EPV.*

#### 3.1.1 Independent variable

We have two independent variables: *EPV* and *Classifier*. Please note that we consider *Classifier* as the main variable of interest and *EPV* as a confounding variable. We use the following set of classifiers since widely used in software engineering studies [38]:

- Random Forest: It generates some separate, randomized decision trees and provides as classification the mode of the classifications. It has proven to be highly accurate and robust against noise [39]. However, it can be highly computationally expensive as it requires the building of several trees.

- Logistic Regression: It estimates the probabilities of the different possible outcomes of a categorically distributed dependent variable, given a set of independent variables. The estimation is performed through the logistic distribution function [40].

- Naïve Bayes: It uses the Bayes theorem, i.e., it assumes that the contribution of an individual feature towards deciding the probability of a particular class is independent of other features in that project instance [41].

- HyperPipes: It simply constructs a hyper-rectangle for each label that records the bounds for each numeric attribute and what values occur for nominal attributes. During the classifier application, the label is chosen by whose hyper-rectangle most contains the instance (i.e., that which has the highest number of attribute values of the test instance fall within the corresponding bounds of the hyper-rectangle)[2].

- IBK: Also known as the k-nearest neighbors' algorithm (k-NN) which is a non-parametric method. The classification is based on the majority vote of its neighbors, with the object being assigned to the class most common among its k nearest neighbors [42].

- IB1: It is a special case of IBK with K = 1, i.e., it uses the closest neighbor [42].

---
[2] https://tinyurl.com/tw9p6zn

- J48: Builds decision trees from a set of training data [43]. It extends the Iterative Dichotomiser 3 classifier [44] by accounting for missing values, decision trees pruning, continuous attribute value ranges and the derivation of rules.

- VFI: Also known as voting feature intervals [45]. A set of feature intervals represents a concept on each feature dimension separately. Afterward, each feature is used by distributing votes among classes. The predicted class is the class receiving the highest vote [45].

- Voted Perceptron: It uses a new perceptron every time an example is wrongly classified, initializing the weights vector with the final weights of the last perceptron. Each perceptron will also be given another weight corresponding to how many examples do they correctly classify before wrongly classifying one, and at the end, the output will be a weighted vote on all perceptron [46].

### 3.1.2 Dependent variable

Our main dependent variable is the accuracy of the classifiers as measured via the area under the receiver operating characteristic curve, i.e., AUC [16]. AUC is created by plotting the true positive rate (i.e., the ratio of classes classified as defective and actually defective over the number of classifications) against the false positive rate (i.e., the ratio of classes classifier as defective and actually not defective over the number of classifications) at various threshold settings. AUC is preferable over other metrics such as F1 because it is threshold independent [33] [47].

To better interpret the AUC, we also observe Precision and Recall which are defined as the true positive divided by positives (e.g., the ratio of classes correctly identified as positives) and the true positive rate, respectively. Further, we also report MCC as an additional performance measure.

### 3.1.3 Analysis procedure

The analysis procedure consists of three main phases: dataset preparation, simulation, and data analysis. Regarding dataset preparation, we developed our datasets by performing a four-step procedure:

1. *Multi-release project creation*: We replicated the project selection performed by Tantithamthavorn et al. [11] and we added two projects of our industrial partner Keymind as they were already successfully used in the past [14]. All of the 101 projects available online[3] used in Tantithamthavorn et al. [11] relates to a single release of a project. Thus, in one project, we merge in the data related to different releases of the same project by keeping the information about the related release. For instance, we create the project Ant by merging projects Ant 1.3, Ant 1.4, Ant 1.5, Ant 1.6 and Ant 1.7 and by adding a column called "Release ID" reporting the release number of a class. We note that Release ID is used for preserving the order but not as a predictor variable.

---

[3] https://tinyurl.com/v8eout6

2. *Mono- and bi-release project removal*: We removed projects that have less than three releases. This led to 16 projects.
3. *Unfeasible project removal*: We remove datasets where a single classifier took more than two weeks for the simulation to end; this led to 12 projects. We set the threshold at two weeks since this is the recommended release duration in agile projects [48] and hence the period at which a model shall be used and hence chosen.

Table 2: Characteristics of the used 15 datasets.

| Dataset | Number of releases | EPV | Number of observations | Number of features |
|---|---|---|---|---|
| ant | 5 | 18 | 1692 | 21 |
| ar | 5 | 2 | 428 | 30 |
| camel | 4 | 28 | 2784 | 21 |
| ivy | 3 | 6 | 704 | 21 |
| jedit | 5 | 15 | 1749 | 21 |
| Keymind-A | 5 | 7 | 702 | 24 |
| Keymind-B | 5 | 3 | 475 | 24 |
| log4j | 3 | 13 | 449 | 21 |
| lucene | 3 | 22 | 782 | 21 |
| poi | 4 | 35 | 1378 | 21 |
| synapse | 3 | 8 | 635 | 21 |
| velocity | 3 | 18 | 639 | 21 |
| xalan | 4 | 90 | 3320 | 21 |
| xerces | 4 | 33 | 1643 | 21 |

Regarding the simulation, Figure 2 explains how we partitioned each dataset. Specifically, for each project with *n* releases (e.g., 4), Part *A* is defined as the first *n-1* releases (i.e., releases 1 to 3), part *B* as release *n* (i.e., release 4). Afterward, we removed the release information, and we trained the classifier on A and tested on B. We implemented the three different techniques in JAVA by using the WEKA API 3.6.15[4]. We used the nine classifiers in the WEKA API. The entire simulation took about two months on a Linux VM, on top of a Cisco UCS C240 M3 Rack Server running VMWare ESXi 5.5 featuring four cores and 16GB of RAM, hosted at Cal Poly. The scripts used for the simulation are available online[3].

| 1 | 2 | 3 | 4 |
|---|---|---|---|
|   | *n-1* |   | *n* |
|   | A |   | B |
|   |   |   |   |
| Release |   |   |   |

Figure 2: A common usage scenario where the classifier is trained on all releases other than the last and is tested on the last.

---
[4] https://sourceforge.net/projects/weka/files/weka-3-6/3.6.15/

To measure the impact of EPV on the classifier accuracy, we have divided the 14 datasets into two groups of equal size according to their EPV; Low or High. This setup minimizes Type II error by maximizing the number of observations for each treatment [49]; therefore, the design minimizes the likelihood to not reject the hypotheses due to a small number of observations. Regarding the data analysis, we compared the extent to which the accuracy varies across each classifier. Specifically, we performed a 2-way ANOVA [50]. In this and the following research question, we use a confidence level, i.e., alpha, of 5% as is standard in software engineering studies.

We use eta squared ($\eta2$) to perform the effect size analysis. We prefer eta squared over other options because our independent variables are categorical (EPV and classifier), it is more conservative than partial eta squared in estimating the size of the effect, and it provides the proportion of variance explained for each factor in an intuitive way controlling for the other factors included in the model (i.e. variance explained sum up to 1 and in case of EPV it is equivalent to correlation coefficient since degrees of freedom for EPV is one).

### 3.2  RQ2: Do techniques vary in accuracy?

In this research question, we investigate if validation techniques differ in the accuracy on estimating model performances. Thus, our null hypotheses are:

*H02a: The accuracy of walk-forward is the same as 10-fold cross-validation.*

*H02b: The accuracy of walk-forward is the same as bootstrap.*

#### 3.2.1  Independent variable

Our independent variable is the validation technique. The time-series technique we use is walk-forward because our literature review concluded it is the only time-series technique used in previous studies. Moreover, this technique maximizes data usage and hence it is particularly effective on small datasets as the ones in our contexts. As a non-time-series technique, we used 10*10-fold cross-validation (hereinafter called 10-fold) and out-of-sample bootstrap (hereinafter called bootstrap) because they are the most used and the only recommended validation techniques, respectively.

Regarding the recommendation aspect of accuracy, analyzing techiques performances in terms of AUC might be misleading since the feasible spectrum of AUC is constrained by the performances of the classifiers that the technique can recommend. In other words, a technique AUC of 1.0 might be unfeasible to reach since no classifier among the ones that the technique can recommend, generally has a perfect accuracy. Similarly, a technique providing an AUC of 0.90 shall be interpreted as poorly accurate in recommending classifiers if most of the classifiers provides a higher AUC than 0.90. Therefore, to facilitate interpretation of AUC results we also consider the following three hypothetical techniques:

1) *Best*: this represents the technique recommending the best classifier always. This technique AUC is interesting since it represents the upper bound of technique's AUC. In other words, the difference between the best real technique accuracy and this best hypothetical technique represents the gain that is potentially achievable in future works.

2) *Medium*: this represents the technique recommending the median classifier always. This technique AUC is interesting since it represents the results of not using any techniques, i.e., using a median classifier among the available ones. In other words, the difference between a real technique's accuracy and this medium hypothetical technique represents the gain in using that technique over not using any techniques.

3) *Worst*: this represents the technique recommending the worst classifier always. This technique AUC is interesting since it represents the lower bound of technique's AUC.

*3.2.2 Dependent variable*

We measure the technique's accuracy in three different aspects:

*Technique AUC*: One aspect of the technique's accuracy is its ability to recommend the most accurate classifier to use. We measure this aspect with the variable called technique AUC, i.e., the accuracy (i.e., AUC) of the classifier that a given technique recommends using. We note that this recommendation aspect relates not only to choosing among classifiers but also among parameters (tuning [4][5][6]) and variables (feature selection [9]) of the classifiers. In other words, a tuned classifier is a classifier using parameters that differ from the default ones. As a matter fact, tuning techniques use validation techniques to measure the accuracy of classifiers adopting specific parameters.

*Bias*: One aspect of the technique's accuracy is the error size, i.e., how different they are to the actual classifier accuracy. We measure this aspect with the variable called absolute bias, as used and defined by previous studies, is the distance between the estimated and the actual classifier accuracy, i.e., |Estimated Accuracy- Actual Accuracy| [11]. We note that we need care in interpreting this metric as it can fail in evaluating the level of support of techniques to recommend classifiers. Specifically, if a technique significantly overestimates the performance of the best classifier, then this technique would have a high error despite recommending the right classifier to use.

*Absolute Bias*: One aspect of the technique's accuracy is the direction of the error. Specifically, previous studies [37] [51] suggest that cross validation overestimates the accuracy of classifiers and hence it is interesting to observe the direction of the error. We measure this aspect with the variable called bias which is defined as Estimated Accuracy- Actual Accuracy. The sign of this variable explains the direction of the error.

We do not measure the stability or variance of techniques because time-series techniques are intrinsically perfectly stable, since they do not use any random mechanism. To the best of our knowledge, this is the first software engineering study comparing techniques on both technique AUC, bias and absolute bias. We do not compare techniques according to threshold dependent metrics such as the ones used in RQ1 (i.e., Precision, Recall and Matthew Correlation Coefficient) since the threshold is context dependent. Moreover, any classifier, even a dummy one, can achieve a perfect Recall just by lowering the threshold. Thus, the classifier selection is based on a threshold independent metric (i.e., AUC) only.

*3.2.3 Analysis procedure*

To measure the *technique AUC* (see Section 3.2.1), we first measure the AUC of each classifier by using each technique on part A of the dataset reserved as the validation set (see Figure 2). Please note that the way different techniques partition A into training and validation sets differ. Afterward, for each technique, we chose the classifier with the highest AUC in the validation set of A; this is the classifier that a given technique suggests using on B for testing. Finally, the technique AUC is the AUC that the suggested classifier provides when trained and validated in A and tested in B (as measured in RQ1). In other words, we refer to a technique's accuracy as the accuracy of the classifier tested in B. Thus, the partitioning of the datasets, as in Figure 2, results in a test set B and a training (and validation) set A, both sets are constant among techniques. Partition A is further divided into training and validation sets depending on how the validation technique uses a set of given data (see Figure 1).

To measure bias, and absolute bias, we simply compare the AUC in A provided by a given technique with the AUC that the classifier provides when trained on A and tested on B.

To analyze results, we apply paired t-test [52] after having checked the data belong to a normal distribution (i.e., we do not have evidence the data does not come from a normal distribution). We apply the Wilcoxon-signed-rank test [53] if one or more distributions result different to a normal distribution. We use Cohen's d to perform the effect size analysis.

### 3.2.4 Sanity check

Suppose we have a dataset related to a project of five releases and that the proportion of defective classes in releases four and five is significantly lower than in the previous three releases. In a realistic context, to predict release four, the classifier is trained on releases one to three; a technique that measures the accuracy of the classifier on release four would bias results if data of release five is used during training. If this is true, then, to provide realistic results, it is important to ensure that train set data precede the test set data. If the order needs to be preserved, then the assumptions of non-time-series techniques are not met and hence they cannot be used; i.e., time-series technique are expected to outperform non-time-series-techniques. Therefore, if the defect rate in later releases of a project is different from the ones in earlier releases, then it is important to ensure that training set data is prior to the test set data. Thus, we want to conduct a sanity check for RQ2: 1) to demonstrate that fluctuations in defect rates may explain any potential difference between time-series and non-time-series techniques 2) and to reinforce the need to use time-series techniques in the within-project across-release class-level context in a way that is independent by the specific experimental choices required by RQ2. Therefore, our null hypothesis for this sanity check is:

*H03: There is no difference in defective rate between the first and second half of a project.*

In this analysis, our independent variable is time as measured by release order. We chose this measurement because current datasets contain classes related to specific releases of a project and releases are ordered (e.g., Ant 1.1 is antecedent Ant 1.2). We note that classes of the same release are not ordered. Our dependent variable is class defectiveness, i.e., if a class has at least one defect or not. We chose this variable because it is exactly what the classifiers predict. We note that this variable is measured for a specific class of a release and it is binary (i.e., True or False).

For conducting this sanity check, we modified the datasets by adding an additional column called "Order" whose value is binary: "first" or "second". Specifically, given a project with *n* releases, all classes of release *m*, where $m > n/2$, are tagged as "second", "first" otherwise. For instance, all classes of releases Ant 1.3, Ant 1.4, and Ant 1.5 are tagged "first", while all classes of releases Ant 1.6 and Ant 1.7 are tagged "second". We opted for this setup in order to minimize the Type II error. Specifically, this setup maximizes the number of observations for each treatment and hence minimizes the likelihood to not reject the hypotheses due to a small number of observations.

Afterward, for each dataset, we compared the defective rate of "first" versus "second" and the distributions of defective rates by performing the Fisher exact test [54]. We chose this test since it is non-parametric and highly recommended when the data is small, i.e., less than 1000 observations, or can be unequally distributed [55]. Since this test makes use of contingency tables, we report odds ratio as a measure of effect size.

# 4 Results

In this section, we present the results for the analyses explained in Section 3, for the two research questions. We provide a detailed discussion of these results later in Section 5.

## 4.1 RQ1: Do classifiers vary in accuracy?

Figure 3 reports the distribution of AUC, Precision and Recall on datasets with Low versus High EPV. Figure 4 reports the same distributions for each classifier of the nine classifiers. According to Figure 3, the difference among classifiers' accuracies are higher in datasets with Low EPV in terms of AUC. Such a result is also perfectly in line with Tantithamthavorn et al. [11]. It is interesting to note that this trend does not apply to every classifier. For instance, according to Figure 4, the classifier IBk and Logistic have a higher difference in accuracy in High versus Low EPV datasets. Moreover, we note that the difference among classifiers' AUC in datasets with Low EPV is still very large (i.e., AUC [0.20, 0.95]) thus suggesting that all datasets should be considered in RQ2.

Moreover, according to Figure 3 and Figure 4, the AUC of classifiers is not higher in datasets with High EPV. Such a result is indeed in contrast with Tantithamthavorn et al. [11] where the higher the EPV, the higher the accuracy. Specifically, in Figure 3 the median AUC is slightly higher than 0.60 in both Low and High EPV cases. Moreover, in Figure 4, we can see that for some classifiers, like HyperPipes and RandomForest, the lower the EPV, the higher the AUC.

It is interesting to see how in both Figure 3 and Figure 4, in both high and low EPV datasets, the distributions of Precision and Recall are much wider than the AUC distributions.

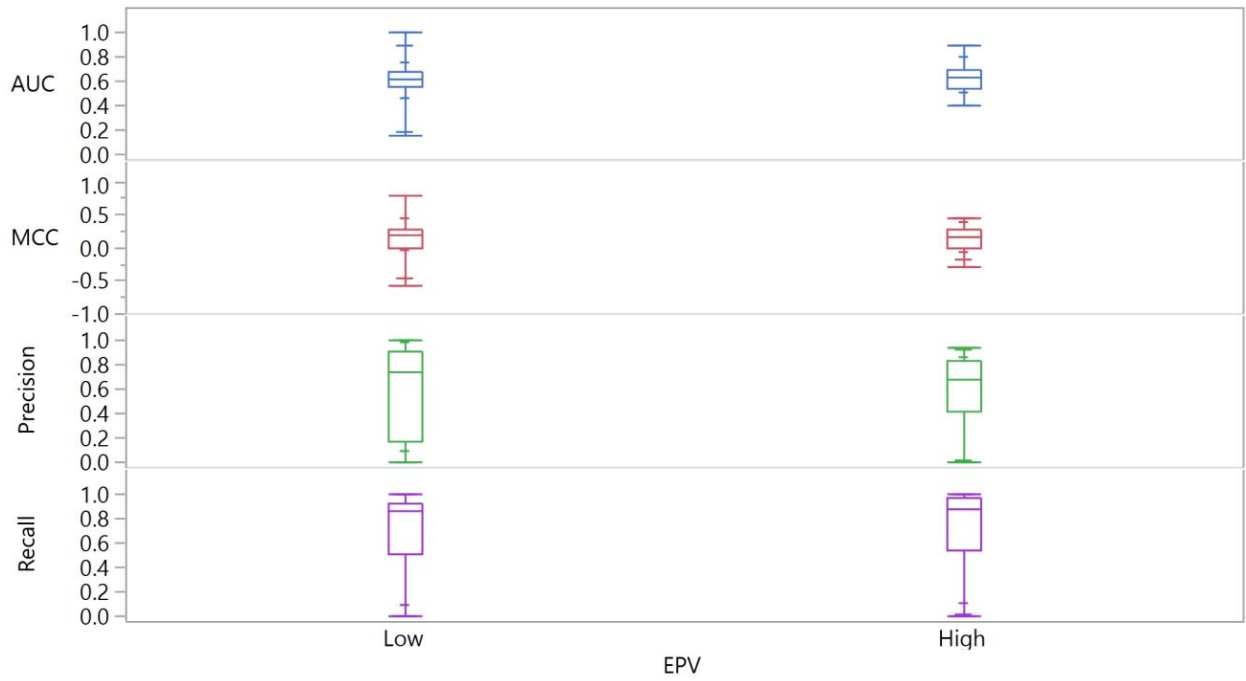

Figure 3: Distribution of AUC, Precision and Recall, of the nine classifiers on datasets with Low versus High EPV.

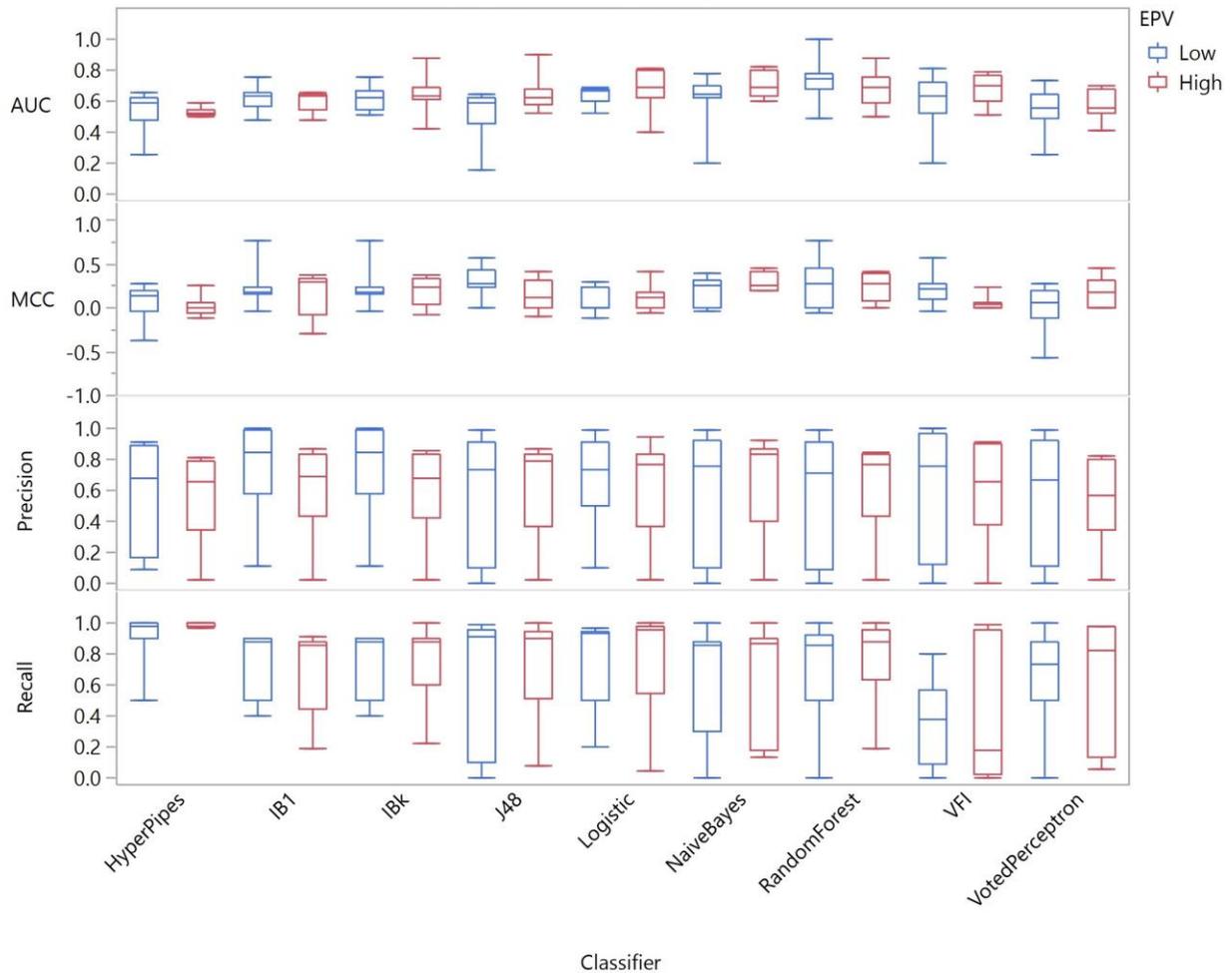

Figure 4: Distribution of AUC, Precision and Recall of each of the nine classifiers on datasets with Low (blue) versus High (red) EPV.

Table 3 reports on the statistical significance test and the effect size of the choice of the classifier and EPV on AUC, MCC, Precision and Recall. Regarding the importance of selecting classifiers, according to Table 3, the classifier does influence and the EPV does not influence, AUC, MCC and Recall. Therefore, we can reject H01. The effect size ($\eta^2$), with respect to classifier choice on AUC, is 0.159 (small). Regarding the effect size analysis, EPV and classifier explains together the 17.8% of variation in AUC. Since we are not after a full explanatory model that best describes AUC, this can be considered reasonable, though total proportion of variance explained is low. EPV is neither a statistically significant factor nor has a meaningful effect in explaining AUC, having less than 2% correlation. Classifier, on the other hand, is statistically significant and has a non-negligible effect, explaining 15.9% of the variation in AUC results (with post-hoc power analysis yielding 94.8%). Choice of classifier is also significant with respect to MCC and Recall with small effect sizes 0.132 and 0.181, respectively. Neither classifier nor EPV is significant with respect to precision of the models, though.

Table 3: Statistical significance and effect sizes for the influence of classifier and EPV on the accuracy metrics.

|  | AUC | | MCC | | Precision | | Recall | |
| --- | --- | --- | --- | --- | --- | --- | --- | --- |
|  | p-value | $\eta^2$ | p-value | $\eta^2$ | p-value | $\eta^2$ | p-value | $\eta^2$ |
| Classifier | **0.007** | **0.159** | **0.032** | **0.132** | 0.976 | 0.018 | **0.003** | **0.181** |
| EPV | 0.106 | 0.019 | 0.477 | 0.004 | 0.795 | 0.001 | 0.785 | 0.001 |

### 4.2 RQ2: Do techniques vary in accuracy?

Regarding the recommendation accuracy of techniques, Figure 5 and Figure 6 report the technique's AUC. According to Figure 6, walk-forward outperforms 10-fold cross-validation and bootstrap in the AUC of the recommended classifier. Specifically, among the three techniques:

1. walk-forward recommends the classifier with the highest average AUC among datasets.

2. walk-forward recommends the classifier with the highest AUC in the highest number of datasets, i.e., both industry projects and 10 out of 12 open-source projects.

Regarding the statistical test, all three real distributions resulted as being not statistically different from a normal distribution. Thus, we proceeded by performing the t-test for comparing the accuracies of the classifiers recommended by the different real techniques. Table 4 reports on the statistical test on the difference in the accuracy of the classifier recommended by the three techniques. According to Table 4 the difference between walk-forward and 10-fold cross-validation and between walk-forward and bootstrap is statistically significant with medium effect sizes. Thus, we can reject both H02a and H02b in terms of AUC.

Regarding susceptibility and inflation, Figure 7 reports the distribution of Bias and Absolute Bias of the three techniques. According to Figure 7, the out-of-sample distribution has the lowest first percentile and the highest last percentile thus suggesting it is the technique providing the highest underestimation and overestimation errors. The walk-forward distribution is much smaller than the distribution of the other two techniques thus suggesting it is the techniques proving a more stable type of error. Since the median (i.e., fiftieth percentile) of the walk-forward distribution is the closest to zero, then walk-forward provides the best compromise between overestimation and underestimation error. Since the median of all three techniques is higher than zero, then all techniques are more prone to overestimate than to underestimate; walk-forward overestimates the least among the techniques.

According to Figure 7, the out-of-sample distribution has the highest median and last percentile thus suggesting it is the technique providing the highest error in the average and in the worst-case scenario. Since the median of the walk-forward distribution is lower than the other two techniques, then the walk-forward provides the smallest error, i.e., walk-forward is the most accurate technique.

Regarding the statistical test on susceptibility and inflation, one or more distributions resulted as being different from the normal one and hence we proceeded by performing the non-parametric Wilcoxon-signed-rank test for comparing the Bias and Absolute Bias of the different techniques. Moreover, the p-value of the statistical test between the AUC of Walk-forward and best is 0.011 and between Walk-forward and medium is less than 0.001. Corresponding effect sizes are small to medium. Thus, we can reject both H02a and H02b in terms of bias and absolute bias.

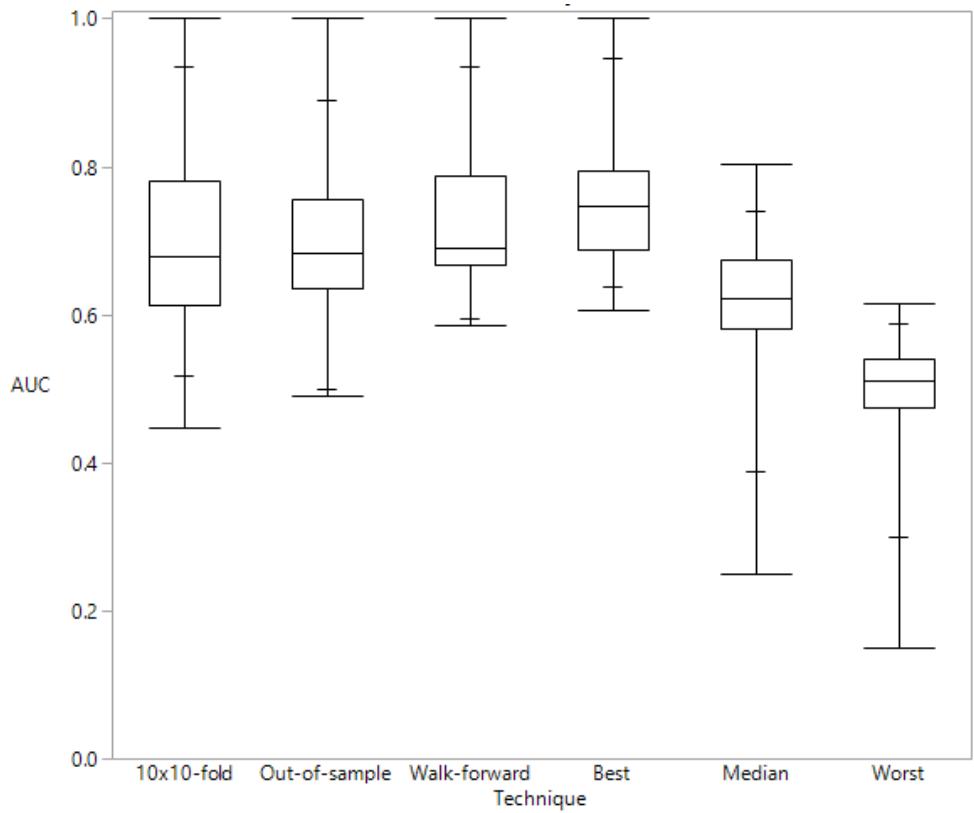

Figure 5: Accuracy of the classifier recommended by a technique (x-axis).

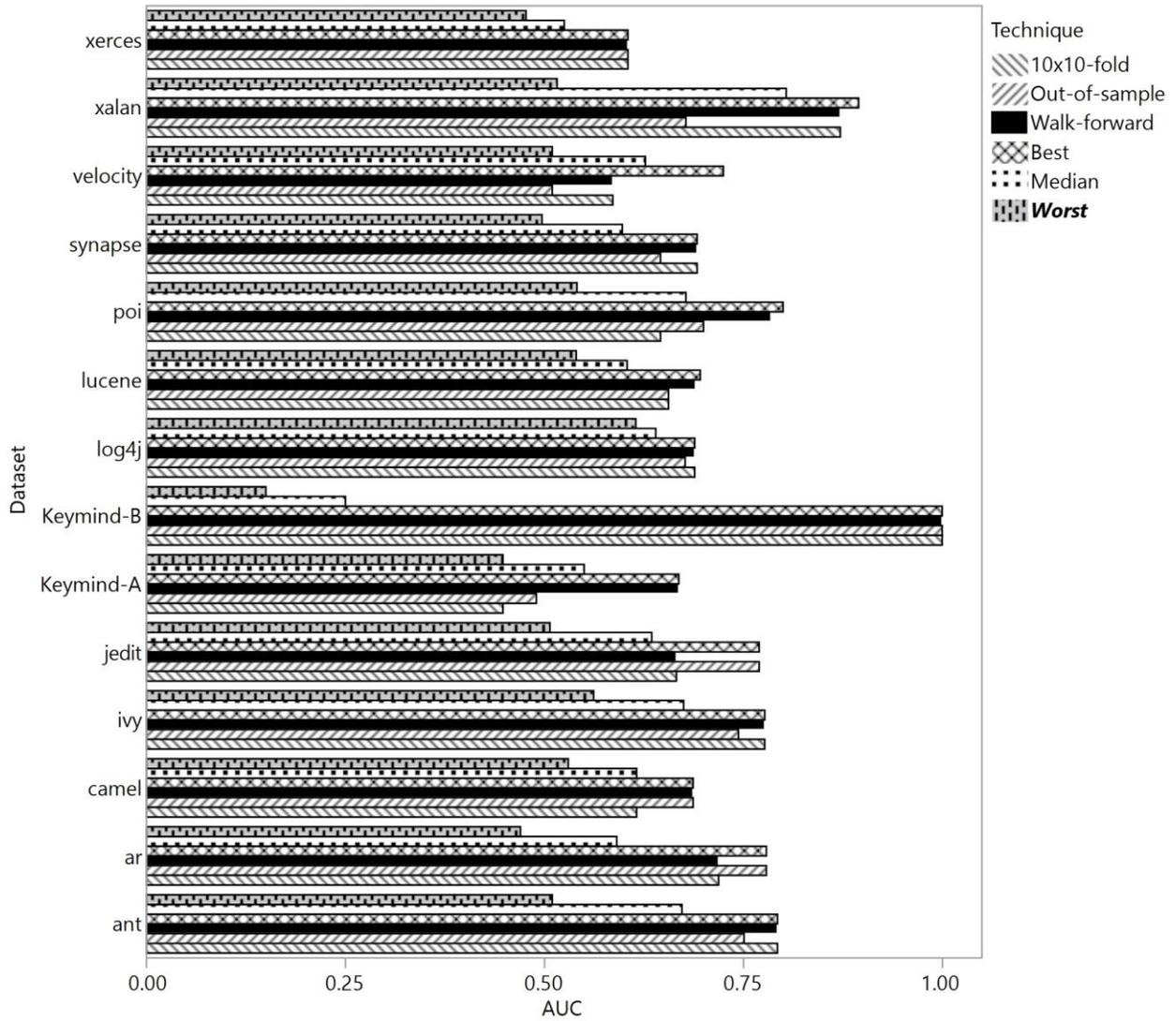

Figure 6: Accuracy of the classifier, in a dataset (x-axis), recommended by a technique (color).

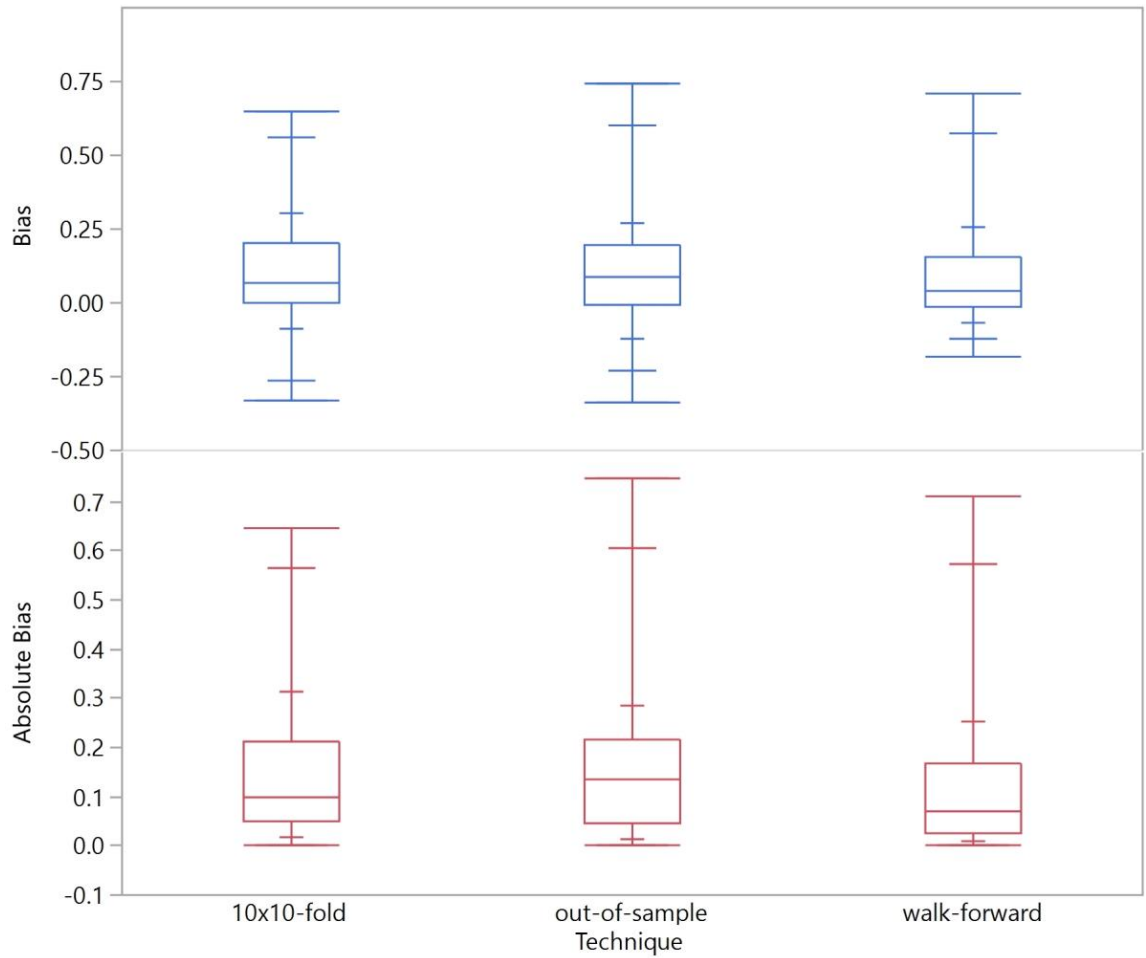

Figure 7: Distribution of Bias and Absolute Bias of the three techniques.

Table 4: Statistical significance and effect sizes for the difference in the accuracy of techniques.

|  | AUC | | Bias | | Absolute Bias | |
|---|---|---|---|---|---|---|
|  | p-value | Cohen's d | p-value | Cohen's d | p-value | Cohen's d |
| Walk-forward vs. 10x10 cv | 0.043 | 0.467 | <<0.001 | 0.220 | << 0.001 | 0.422 |
| Walk-forward vs. Out-of-sample | 0.047 | 0.509 | 0.011 | 0.136 | << 0.001 | 0.327 |

*4.2.1 Sanity check*

Table 5 reports the defective rates of the first and second half of a dataset. According to Table 5, the relative difference between the defective rate of the second and first half of a project is in the range [-75%, 218%]. Since the distributions of defective rates between the first and the second half are statistically different in 12 out of 14 datasets, we can reject H03 in both industry projects and 83% of open-source projects.

Table 5: Defective rate in the first and second halves of ordered data in the dataset. The average defective rate of the two Keymind projects is not reported due to a non-disclosure agreement.

| Dataset | Defective Rate First Half | Defective Rate Second Half | Difference of Means of Defective Rate | Relative Difference | Pvalue | Effect Size (Odds Ratio) |
|---|---|---|---|---|---|---|
| Ant | 0.154 | 0.235 | 0.081 | 52% | <0.001 | 1.68 |
| Ar | 0.127 | 0.168 | 0.041 | 32% | 0.275 | 1.40 |
| Camel | 0.242 | 0.181 | -0.061 | -25% | 0.002 | 0.69 |
| Ivy | 0.224 | 0.114 | -0.111 | -49% | <0.001 | 0.44 |
| JEdit | 0.274 | 0.069 | -0.206 | -75% | <0.001 | 0.20 |
| Keymind - B | - | - | - | 218% | <0.001 | - |
| Keymind - A | - | - | - | -40% | 0.020 | - |
| Log4J | 0.291 | 0.922 | 0.631 | 217% | <0.001 | 28.13 |
| Lucene | 0.532 | 0.597 | 0.065 | 12% | 0.083 | 1.30 |
| Poi | 0.323 | 0.640 | 0.317 | 98% | <0.001 | 3.72 |
| Synapse | 0.201 | 0.336 | 0.135 | 68% | <0.001 | 2.01 |
| Velocity | 0.705 | 0.341 | -0.364 | -52% | <0.001 | 0.22 |
| Xalan | 0.326 | 0.730 | 0.404 | 124% | <0.001 | 5.58 |
| Xerces | 0.246 | 0.486 | 0.240 | 98% | <0.001 | 2.90 |

# 5  Discussion

Our analysis of the literature show that only 9% of studies used a time-series technique (see Table 1); this means that the great majority of past studies aimed at estimating the performance of a classifier on temporally random data rather than measuring it on future data. However, our experience in talking with practitioners during technology transfer of classifiers [25] shows that it makes more sense to find evidence on a statement like "This technology would have helped us if we had used in the past" (i.e., measuring on data to have come) than evidence for a statement like "This technology can help us if we use it in the future" (i.e. estimating on temporally random data). In this context, it is especially important to consider we have no better way to make assumptions about future data from past data. To make an analogy with human-based experiments for validating software engineering technologies via controlled experiments, we look at past performances of the subjects using different treatments to generalize into future performance of the sampling population [26]. The rationale is to allow the potential technology user to interpret the results and decide how much the future is different from the past and hence how the experimental results are generalizable to the specific usage context. Regarding the comparison among validation techniques, here we discuss the main results, takeaways and implications from both of our research questions.

Our results suggest that EPV does not have any statistically significant effect on any of the three performance measures, but the choice of classifier has a statistically significant effect on performance in terms of AUC and recall. In other words, classifiers do not statistically vary in terms of Precision; this is confirmed in Figure 4 where the difference among classifiers is less notable in Precision than in AUC, MCC or Recall. Thus, a good technique would support you in using a classifier providing a better overall accuracy (AUC and MCC) and in the number of identified defective classes (Recall) but not in the ratio of classes correctly identified as defective. Thus, the main takeaway from RQ1 is that *it is important to carefully choose classifiers* and this motivates RQ2.

The comparison of AUC of the classifier recommended by walk-forward versus the median classifiers shows an average improvement of 33% among datasets. Thus, choosing the classifier via walk-forward increases accuracy by 33% when compared to selecting the classifier by not using any technique. Moreover, walk-forward is the only technique that recommends the classifier with an AUC higher than the median classifier in both industry and all but one open-source project. For instance, in Keymind-A walk-forward is the only technique that recommends the classifier with an AUC higher than the median classifier. It is interesting to note the case of Velocity where none of the three techniques outperforms the use of no technique.

It is interesting to note that walk-forward, i.e., the best real technique, resulted statistically worse than the best hypothetical technique and statistically better than the median hypothetical technique. This means that the use of walk-forward is highly recommended over using no technique or other non-time series technique. However, since there is a significative difference with the best hypothetical technique, then this is a promising area of improvement for future works.

Regarding the overestimation of bias observed in Figure 7, the fact that cross-validation overestimates the accuracy of classifiers is perfectly in line with previous studies [37] [51] [56] suggesting that. However, no previous study shows that also walk-forward overestimates.

In conclusion, the main takeaway from RQ2 is that *walk-forward resulted significantly more accurate in choosing accurate classifiers than both 10-fold cross-validation and bootstrap*.

It is interesting to note that the relation between time and class defectiveness is not monotone, i.e., classes do not constantly tend to have a higher or a lower defective rate over time. Specifically, in eight cases (out of 14 projects) the rate is higher in the second half, and in the remaining 6 cases, it is higher in the first half. This result also applies to the two industrial datasets; in Keymind-A, the higher defective rate is in the first half, in Keymind-B it is in the second half. Anyway, regardless of the sign, the average relative difference, among datasets, is very high: 82%. This high difference is not considered by the sampling procedure adopted in non-time-series techniques which makes validation results poorly realistic.

As intuitive, according to Table 2 and Table 5, a high p-value was the result of a low relative difference among the halves of a project and a small number of observations in the dataset as we intuitively expected. For instance, if we compare datasets Ar with Camel, Ar has a has a much higher p-value (0.300 > 0.002) since it has a slightly higher absolute relative difference (32% > 25%) and a much lower dataset size (428 < 2784).

Considering RQ2 results and the analysis for sanity check, we note that in most of the datasets 1) the walk-forward significantly outperforms the other two techniques in the accuracy of the recommended classifier (see Figure 6) and 2) the defective rate of a project statically changes over time (see Table 5). It is interesting to note the case of Ar which is one of the two datasets where the defective rate is not statistically different and is also one of the two datasets where bootstrap outperforms walk-forward. Thus, results seem to indicate that the change in the defective rate of a project is the reason why walk-forward outperforms the other two techniques in the accuracy of the recommended classifier. JEdit is the only exception, in fact, in JEdit bootstrap outperforms walk-forward in the accuracy of the recommended classifier despite the defective rate statically changing over time. However, we recommend care in correlating datasets characteristics to validation technique accuracy since a valid analysis needs a high number of heterogeneous datasets.

In conclusion, ***the data managed by the classifier in the within-project across-release class-level context requires the technique to preserve the order of data***. Thus, walk-forward is not only more accurate than 10-fold cross-validation and bootstrap (RQ2) but is also the only one that respects the property of the data, i.e., it preserves the order of data.

## 6 Threats to validity

In this section, we discuss possible threats to validity related to our study. The threats are organized by type (i.e., Conclusion, Internal, Construct, and External).

Conclusion validity regards issues that affect the ability to draw accurate conclusions about relations between the treatments and the outcome of an experiment [57]. In both research questions we used non-parametric tests; thus, we are more prone to Type I errors (i.e., rejecting a true null hypothesis) than to Type II errors (i.e., non-rejecting a false null hypothesis). This is particularly relevant for RQ2 where the ability to reject the hypotheses has also been inhibited by the very low number of data points, i.e., 14. Thus, we recommend care in judging as not significant the difference in the accuracy of the classifier recommended by the different techniques. Another important threat to construct validity is measuring a classifier's accuracy by using a single dependent variable, i.e., AUC. This choice was driven by previous researchers recommending to avoid threshold-dependent metrics such as Precision and Recall and to use AUC [33]. We did not use additional threshold-independent metrics such as the Brier score [58] mainly because they were practically irrelevant for our context: i.e., choosing a classifier. Finally, to support the replicability of data analysis, we report our raw results online[5].

Internal validity regards the influences that can affect the independent variables concerning causality [57]. The only threat to the validity of this type that is relevant in this study is the use of halves as the measure of temporal order. An alternative approach could have been using the release ID as a metric of temporal order. However, this alternative approach would have reduced the number of observations per treatment. This reduction would have increased the risk of Type

---

[5] https://tinyurl.com/uyge98x

II error, which is already high due to the use of non-parametric-tests (see above discussion), without reducing the risk of Type I error.

Construct validity regards the ability to generalize the results of an experiment to the theory behind the experiment [57]. This threat to validity is low as the used datasets have been already successfully used and published by other researchers; we did not add any information to the datasets. Since this study is about how to choose classifiers and hence on how to validate technologies aimed at improving classifiers accuracy, then we did not used any technologies such as noise removal [1][2][3], tuning [4][5][6], rebalancing [7][8], and feature selection [9]. In other words, using such technologies might have provided concepts drift [59].

One possible threat to validity is in the specific classifiers used. Despite not considering classifiers that are boosting-based or Neural Network-based, our classifiers are heterogeneous and high in number when compared to similar past studies [11] [36] especially because in RQ2 we focused on their selection. In other words, the findings could change with a different or larger set of classifiers. However, the more is not the better. It is a good practice to analyze classifiers that we expect to perform well; any classifier, even a random one, can be accurate randomly. Thus, we based our classifier selection according to related works. Finally, we note that the number of releases in our projects, including the industrial ones, is low, specifically between three and five. Thus, we recommend caution in generalizing these results in contexts with a high number of releases.

External validity regards the extent to which the research elements (e.g., subjects and artifacts) are representative of actual elements [57]. To mitigate this threat, we used all multi-release datasets we knew are publicly available. We note that all the used projects, including the two industrial ones, have a low number of releases and hence it is the context of our study. Projects with a high number of release must be analyzed by a completely different approach such as moving window [36] [60].

Finally, to promote replicability, we made our scripts and datasets available online[3].

# 7 Conclusions

This paper reflects on the importance of preserving the order of data when validating classifiers, i.e., in the use of the time-series technique. We use the last release of a project as the ground truth to evaluate the classifier accuracy and hence the ability of a technique to recommend an accurate classifier. We consider nine classifiers, two industry and 13 open projects, and three validation techniques: namely 10-fold cross-validation (i.e., the most used technique), bootstrap (i.e., the recommended technique), and walk-forward (i.e., a technique preserving the order of data).

Our results show that it is important to choose a classifier by using a technique since their accuracy significantly vary regardless of the dataset EPV. Moreover, the relative difference in defect rate between the second and first half of a project is, among projects, in the range [-75%, 218%] and it is statistically different in both industry and 10 out of 14 open-source projects. Since non-time-series techniques randomly sample the datasets to create test and training sets, they do not preserve such differences which exist in practice. Therefore, the accuracy measured by non-time-series techniques is poorly realistic of any classifiers' practical adoptions. This is reflected by the fact that walk-forward (a time series technique) outperformed both the other two techniques statistically in all three accuracy metrics: AUC of the selected classifier, bias and absolute bias. Surprisingly, all the techniques resulted to be more prone to overestimate than to underestimate the performances of classifiers; however, walk-forward overestimates the least among the techniques. Thus, in addition to being by nature more simple, inexpensive, and stable, according to our empirical study walk-forward should be preferred over the other two techniques. Several studies recommend preserving order of data [61][62][63][64] but time-series techniques are not widely employed in practice (see Section 2.2). This paper contributes to the body of work in defect prediction about preserving the order of data for validation and continue raising an alarm to our community.

If on the one side, we recommend in general validating classifiers by using time-series techniques because they are more accurate, simpler, faster and more stable, on the other side, we need to keep in mind there is no silver bullet in software engineering [65]. Similarly, our take is that when choosing the technique to use we must carefully consider the classifier usage scenario, the type of research question, the conclusions to draw, and, when possible, validating the techniques empirically (i.e., meta-validation). As different types of techniques measure different types of accuracies; no technique can be claimed better or worse than another overall. Time-series techniques have both advantages and disadvantages when compared to non-time-series techniques. One of the main advantages is that they replicate a realistic usage scenario [13][14][6]. The rationale is that learning from the future is unrealistic [11][23]. Specifically, our experience in industrial contexts [13][14][6] shows that data of different releases are collected over time, and this data is used to predict defects of the next release of the same or different project. However, the replicated scenario is not universal as the rate at which a classifier is refreshed varies across contexts, i.e., not all past releases could be available to predict the next one. Another advantage is their ability to be relatively inexpensive and fast as the number of runs is equivalent to the number of ordered parts, which in the 101 datasets used by Tantithamthavorn et al. [11] is on average five (releases). Another important advantage is that they are not affected by any bias related to the randomness with which the training and test sets are generated [66][32]. An important disadvantage is that they require more than one set of (ordered) data; in software engineering terms this means multiple releases of the same project. Specifically, we have manually analyzed the 101 datasets used by Tantithamthavorn et al. [11], and we found that 91 projects are multi-release. For instance, the project Ant has five releases whereas the project Tomcat has only one release; thus, only non-time-series techniques can be used on Tomcat.

Our results show that there is a promising area of improvement for future works on choosing the classifier to use. Specifically, the classifier chosen by walk-forward provided an AUC that is statistically lower than the best classifier. This supports future efforts in developing better validation techniques. Other software engineering studies used timeseries techniques different from walk-forward. For instance, McIntosh and Kamei [36] use both of long-period and short period training models, whereas we consider only long-period period models. Moreover, Tan et al. [51] introduce the concept "gap" as time-series techniques. Thus, future studies should evaluate and compare walk-forward with other time-series techniques and short period models.

In the future, we plan to replicate previous studies that validated technologies using bootstrap or 10-fold by using walk-forward to check if the results still hold. Moreover, we plan to compare the use of single techniques versus their combination. The rationale is that no single technique is perfect, the different techniques have pros and cons, and they complement each other. Therefore, combining the different techniques might significantly increase the ranking accuracy and decrease the error bias over the use of any single technique. Finally, since the execution of some techniques on some releases took more than two weeks, then we plan to validate the use of lighter validation techniques.